\documentstyle[aps,epsf,pra,multicol,psfig]{revtex}

\begin{document}

\title{Transient properties of modified reservoir-induced transparency}

\author{D. G. Angelakis, E. Paspalakis and P. L. Knight}

\address{Optics Section, Blackett Laboratory, 
Imperial College, London SW7 2BZ, United Kingdom}

\date{\today}

\maketitle

\begin{abstract}
We investigate the transient response of a $\Lambda$-type
system with one transition decaying to a modified radiation reservoir 
with an inverse square-root singular
density of  modes at threshold, under conditions of
transparency. We calculate the time evolution
of the linear susceptibility for the probe laser field
and show that, depending on the strength of the 
coupling to the modified vacuum and the background decay, the
probe transmission can exhibit behaviour ranging from underdamped
to overdamped oscillations. 
Transient gain without population inversion 
is also possible depending on the 
system's parameters.     
\end{abstract}

\begin{multicols}{2}
It has been now well documented that quantum coherence and interference effects
can modify the absorption and dispersion properties of an atomic system
\cite{Scullybook,Marangos98a}.  
In the most common
situation, that of a $\Lambda$-type three-level system, the medium becomes transparent to a probe laser field near an 
otherwise absorbing resonant transition. This is achieved via the 
application of a second laser field coupling to the linked transition.
In addition to steady state studies, considerable work has been done on the
transient properties of coherent phenomena such as, for
example, electromagnetically
induced transparency \cite{Li95a,Durrant98a}, gain (or lasing)
without inversion \cite{Harris89b,Fry93a,Zhu97b} and coherent population trapping \cite{Knight90a,Korsunsky97a}.

As has been recently shown \cite{Paspalakis99PRAr},
transparency can  occur in the steady state absorption
of a $\Lambda$-type system when one of the
atomic transitions is coupled to a modified 
radiation reservoir having a
threshold with an
inverse square-root dependence of the density of modes,
$\rho(\omega) = \Theta(\omega-\omega_{g})/\left(\pi \sqrt{\omega-\omega_{g}}\right)$, with $\Theta$ being the 
Heaviside step function and
$\omega_{g}$ being the gap frequency. 
Such a density of modes 
can be found near thresholds in waveguides \cite{Kleppner81a,Lewenstein88a}, 
in microcavities \cite{Lewenstein88b,Rippin96a},
and near the edge of a photonic band gap material which is described 
by an isotropic model \cite{John94a,Kurizki94a,Zhu97a,Bay97a,Paspalakis99b}. 
We also note that there is current interest in
coherent phenomena which occur in modified reservoirs
having relatively weak modal densities where the 
Born and Markov approximations can be applied \cite{Kocharovskaya95a,Keitel99a}.

It is known that coherence effects can take 
a considerable time to be set up \cite{Dalton82a},
and the purpose of the present work is to investigate
this question when structured radiation reservoirs are
employed. In this article we study the 
${\it transient}$ behaviour of the absorption
of a $\Lambda$-type system, similar to the one used in 
ref.\ \cite{Paspalakis99PRAr}, where  transparency in the ${\it steady}$ ${\it state}$ absorption spectrum of the system was  predicted. 
In our system, one of the atomic transitions  is
spontaneously coupled to
a frequency-dependent reservoir which displays the above mentioned 
inverse square-root behaviour in its density of modes. Solving the equation of
motion for all times, we show that the rate 
at which the atomic medium becomes 
transparent to the probe field depends crucially on both the background decay rate of the upper atomic level  
and the strength of the coupling to the modified vacuum modes. 
We also find that, under certain 
conditions,  the system can exhibit transient 
gain without inversion. 

The atomic system under consideration is shown in figure 1. 
It consists of three atomic levels in a $\Lambda$-type configuration.
The atom is assumed to be initially in state $|0\rangle$.
The transition $|1\rangle \leftrightarrow |2\rangle$ is taken 
to be near-resonant with a frequency-dependent reservoir, while
the transition $|0\rangle \leftrightarrow |1\rangle$ is assumed to
be far away from the gap and is treated as a free space transition. 
The dynamics of the system can be described using a probability amplitude
approach. The Hamiltonian of the system, in the interaction picture
and the rotating wave approximation, is given by
(we use units such that $\hbar = 1$),
\begin{eqnarray} 
H &=& \bigg[ \Omega e^{i \delta t} |0\rangle \langle 1| + \sum_{{\bf k},\lambda}g_{{\bf k},\lambda}e^{-i(\omega_{\bf k}- \omega_{12})t} |1\rangle \langle2| a_{{\bf k},\lambda} \nonumber \\
&+& \mbox{H.c.} \bigg]
-i \frac{\gamma}{2} |1\rangle \langle1| \, . \label{Ham}
\end{eqnarray}
Here, $\Omega = - $\boldmath${\mu}$\unboldmath $_{01}\cdot$\boldmath${\epsilon}$\unboldmath $E$ 
is the Rabi frequency, with \boldmath${\mu}$\unboldmath $_{nm}$
being the dipole matrix element of the $|n\rangle \leftrightarrow |m\rangle$
transition. The unit polarization vector and the electric field amplitude of the probe laser field are denoted by \boldmath${\epsilon}\,$\unboldmath and $E$ respectively. Also, $\delta = \omega
- \omega_{10}$ is the laser detuning from 
resonance with the $|0\rangle \leftrightarrow |1\rangle$ transition, where
$\omega_{nm} = \omega_{n} - \omega_{m}$ and
$\omega_{n}$ is the energy of state $|n\rangle$ and $\omega$ is the probe
laser field angular frequency. In addition, $\gamma$ denotes the background decay to all other states of the atom. It is assumed that these states are situated far from the gap so that such  background decay can be treated as a Markovian process. 
We note that we are interested in the perturbative behaviour 
of the system to the probe laser pulse, therefore  $\gamma$ can also 
account for the radiative decay of state $|1\rangle$ to state $|0\rangle$
Finally, $g_{{\bf k},
\lambda} = - i \sqrt{2 \pi \omega_{\bf k}/V}$ \boldmath$\epsilon$\unboldmath $_{{\bf k},
\lambda}\cdot$\boldmath$\mu$\unboldmath$_{12}$  where 
$V$ is the quantization volume,
\boldmath$\epsilon$\unboldmath$_{{\bf k},\lambda}$ is the 
unit polarization
vector, $a_{{\bf k},\lambda}$ is the photon annihilation operator and
$\omega_{\bf k}$ is the
angular frequency of the $\{{\bf k},\lambda\}$ mode of the 
modified radiation reservoir vacuum field. 

The wavefunction of the system, at a specific time $t$, can be expanded
in terms of the
`bare' eigenvectors such that
\begin{eqnarray}
 |\psi(t)\rangle &=& b_{0}(t)|0,\{0\}\rangle + b_{1}(t) e^{-i \delta t} |1,\{0\} \rangle \nonumber \\
 &+& \sum_{{\bf k},\lambda}b_{{\bf k},\lambda}(t)|2,\{{\bf k},\lambda\}\rangle \, , \label{wav}
\end{eqnarray}
and $b_{0}(t=0)=1$, $b_{1}(t=0)=0$, $b_{{\bf k},\lambda}(t=0)=0$.
We substitute Eqs.\ (\ref{Ham}) and (\ref{wav}) into the time-dependent Schr\"{o}dinger equation and obtain the time evolution of the 
probability amplitudes as
\begin{eqnarray}
i\dot{b}_{0}(t) &=& \Omega b_{1}(t) \label{a0} \, ,\\
i\dot{b}_{1}(t) &=& \Omega b_{0}(t) - \left(\delta + i \frac{\gamma}{2}\right)b_{1}(t) \nonumber \\
&-& i \int^{t}_{0}dt^{\prime}K(t-t^{\prime})b_{1}(t^{\prime}) \, , \label{a1} 
\\
i\dot{b}_{{\bf k},\lambda}(t) &=& g_{{\bf k},\lambda}e^{i(\omega_{\bf k}- \omega_{12} - \delta)t} b_{1}(t) \, ,
\end{eqnarray}
with the kernel
\begin{eqnarray}
K(t-t^{\prime}) &=& \sum_{{\bf k},\lambda}  g^{2}_{{\bf k},\lambda} e^{-i(\omega_{\bf k}- 
\omega_{12}-\delta) (t-t^{\prime})} \nonumber  \\
&\approx&  \beta^{3/2} \int d\omega \rho(\omega) e^{-i(\omega- 
\omega_{12}-\delta) (t-t^{\prime})} \, ,
 \label{kerneltot}
\end{eqnarray}
and $\beta$ being the atom-modified reservoir
resonant coupling constant. All the coupling  constants ($g_{{\bf k},\lambda}$, $\beta$, $\Omega$) are assumed to be real, for simplicity.

The time evolution of the 
absorption and dispersion properties of the system 
are determined by, respectively, the imaginary and real parts
of the time-dependent linear susceptibility $\chi(t)$. 
In our case, the
susceptibility can be expressed as
\cite{Li95a} 
\begin{equation}
\chi(t) = - \frac{4\pi {\cal N} |\mbox{\boldmath${\mu}$\unboldmath $_{01}$}|^{2}}{\Omega(z,t)}b_{0}(t)b^{*}_{1}(t) \, ,\label{suscep}
\end{equation}
with ${\cal N}$ being the atomic density.
The solution of Eqs. (\ref{a0}) and (\ref{a1}) 
is obtained by means of time-dependent perturbation theory \cite{Harris89b,Paspalakis99PRAr}. 
We assume that the laser-atom interaction is
very weak $(\Omega \ll \beta, \gamma)$ so that $b_{0}(t) \approx
1$ for all times. Then, Eqs. (\ref{a0}) and (\ref{a1}) 
reduce to
\begin{eqnarray}
i\dot{b}_{1}(t) &\approx& \Omega  - \left(\delta + i \frac{\gamma}{2}\right)b_{1}(t) - i \int^{t}_{0}dt^{\prime}K(t-t^{\prime})b_{1}(t^{\prime}) \, . \label{a1a} 
\end{eqnarray}
We further assume that $\Omega(z,t)$ is approximately 
constant in the medium
and with the use of the 
Laplace transform we obtain from Eq.\ (\ref{a1a})
\begin{equation}
{\tilde b}_{1}(s) = \frac{\Omega}{s\left[\delta +i\gamma/2 + i\tilde{K}(s) + is \right]} \, , \label{A1}
\end{equation}
where ${\tilde b}_{1}(s) = \int^{\infty}_{0}e^{-st} b_{1}(t) dt$, 
$\tilde{K}(s) = \int^{\infty}_{0}e^{-st}K(t)dt$. 
The amplitude  $b_{1}(t)$ is given by the inverse Laplace transform
\begin{equation}
b_{1}(t)=\frac{1}{2\pi i} \int_{\epsilon-i \infty}^{\epsilon+i \infty}
 e^{st} {\tilde b}_{1}(s) ds \, ,
\end{equation}
where  $\epsilon$ is a real number chosen so that $s=\epsilon$ 
lies to the right 
of all the singularities (poles and branch cut points) of function ${\tilde b}_{1}(s)$. 

For the case of an inverse square-root singularity in the
frequency-dependent reservoir density of modes $\tilde{K}(s) = \beta^{3/2} e^{-i \pi/4}/\sqrt{s + i(\delta_{g} - \delta)}$ with $\delta_{g} = \omega_{g} - \omega_{12}$, 
the inverse Laplace transform of Eq. (\ref{A1}) yields
\begin{eqnarray}
b_{1}(t)&=& \sum_{i=1}^{5}\alpha_{i} (x_{i}+y_{i})e^{{x_{i}^2}t} \nonumber \\
&-& \sum_{i=1}^{5}\alpha_{i}y_{i}\left[1-\mbox{erf}(\sqrt{x_{i}^{2}t})\right]
e^{{x_{i}^2}t}, \label{a1t}
\end{eqnarray}
where $y_{i}=\sqrt{x_{i}^2}$ and $x_{i}$ are the roots
of the equation 
\begin{equation}
x^5+c_{3}x^{3}+c_{2}x^{2}+c_{1}x+c_{0} = 0 \, .
\end{equation}
Here
$c_{3}=\gamma/2 -i (\delta_{g}+\delta^{\prime})$, $c_{2}=-iK_{0}$, 
$c_{1}=-\delta^{\prime}(\delta_{g}+i\gamma/2)$, 
$c_{0}=-K_{0}\delta^{\prime}$, $\delta^{\prime}=\delta_{g}-\delta$,
$K_{0}=i \beta^{3/2} e^{-i\pi/4}$ and  $\mbox{erf}$ is the error function \cite{bookintegrals}. The roots of this equation are determined numerically.
The expansion coefficients $\alpha_{i}$ are given by 
\begin{eqnarray}
\alpha_{i}=\frac{i\Omega x_{i}} {(x_{i}-x_{j})(x_{i}-x_{k})(x_{i}-x_{l}) 
(x_{i}-x_{m})}, 
\end{eqnarray}
with $i$,$j$,$k$,$l$,$m$
$=1,2,3,4,5$. Also, if $\mbox{Re}(x_{i})>0$ we have
$y_{i}=x_{i}$, while if $\mbox{Re}(x_{i})<0$ we have $y_{i}=-x_{i}$, 
in order to keep the phase angle of $x_{i}^2$ between $-\pi$ and $\pi$ \cite{bookintegrals}. In addition, if $x_{i}  = 0$
then $\alpha_{i}=0$. Therefore, at least two roots and at most
three roots contribute on the solution (\ref{a1t}) depending
on the system parameters.

Within our perturbative approach, Eq. (\ref{suscep})
yields $\chi(t) \sim - b_{1}^{*}(t)$, 
where $b_{1}(t)$ is given by Eq. (\ref{a1t}).
As has been shown in ref.\ \cite{Paspalakis99PRAr},
steady state transparency occurs for the case that $\delta = \delta_{g}$.
This is the case that also interests us here.
In figure 2 we plot the time evolution of the
imaginary part of the linear, time-dependent susceptibility
$\left[-\mbox{Im}\left(\chi(t)\right)\right]$ 
for different values of the
background decay $\gamma$ and with $\delta=\delta_{g}=0$. 
In the case that $\gamma \gg \beta$,  the susceptibility is always
positive (which denotes absorption in our
convention), has a maximum and the steady state value
is reached adiabatically. If $\gamma = \beta$, the system exhibits only 
absorption, however small oscillations are visible at the beginning.
As $\gamma$ decreases, then these oscillations become more
pronounced, and small gain (or lasing) without the presence of population inversion
between $|1\rangle$ and $|0\rangle$, shown by negative values
of the 
time-dependent linear susceptibility,
is found. If the background decay decreases further and reaches
the regime that $\gamma \ll \beta$ the oscillations increase 
further, the gain without inversion increases, the interaction
becomes more non-adiabatic and the steady state value 
is reached for very large times.  

The behaviour displayed in the previous figure can be
understood if the time evolution of
the population of the excited state $|1\rangle$ is examined. 
As can be
seen from figure 3, after an initial weak absorption 
the population of the state $|1\rangle$ can either decay
smoothly to zero (for the case  $\gamma \gg \beta$) or
evolve by undergoing damped Rabi oscillations between states
$|1\rangle$ and $|2\rangle$ due to reversible
decay which arises via the interaction with the modified 
reservoir \cite{John94a,Kurizki94a}.
These oscillations increase as 
the background decay decreases compared to
the coupling strength to the frequency-dependent radiation reservoir.
In such a way a time-dependent coherence between states
$|1\rangle$ and $|2\rangle$ is created which is responsible
for the phenomenon of transient gain without inversion
shown in figure 2.

This behaviour of the system is related to the 
one predicted \cite{Li95a} and experimentally 
observed \cite{Durrant98a} in a typical 
three level $\Lambda$-type atomic system which exhibits
electromagnetically induced transparency through the application of
a coupling laser field. The difference in our case, is that
the transparency and the transient gain without
inversion occur due to the coupling to a radiation reservoir
with an inverse square-root singularity of the
density of modes at threshold and are not induced by an
external laser field.

In summary, we have discussed the transient 
properties of the transparency
in a $\Lambda$-type atom in which one 
transition spontaneously decays 
to a specific frequency-dependent radiation reservoir. 
The time evolution of the absorption and
thus the way that the steady state is reached depends  crucially
on the background decay rate and the strength of the coupling to the 
modified reservoir.
Transient gain without population inversion is found to exist
if the coupling strength
to the modified reservoir is larger than the background decay rate.
We have only been concerned with the time evolution
of the linear absorption properties of the medium. 
The time evolution of the dispersive
properties of the system, which is another topic of interest
\cite{Paspalakis99PRAr}, will be discussed 
separately. In such a study the simple relationship
between the real part of the susceptibility
and the group velocity cannot be applied (as it holds
only for the steady state), and a different approach
needs to be implemented.

\section*{Acknowledgments}
E.P. thanks Niels Kylstra for helpful
discussions in the subject. We would like 
to acknowledge the financial
support of the UK Engineering and 
Physical Sciences Research Council (EPSRC), the 
Hellenic State Scholarship Foundation (SSF) 
and the European Commission TMR Network on Microlasers and Cavity QED.

\end{multicols}

\begin{figure}
\centerline{\psfig{figure=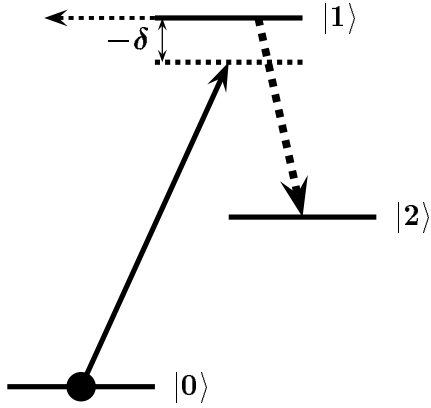,width=4.5cm}}
\caption{\narrowtext  
The system under consideration. The solid 
line denotes the probe laser coupling, the 
thick dashed line
denotes the coupling to the modified radiation reservoir and
finally the thin dashed line denotes the background
decay.}
\label{fig1}
\end{figure}


\begin{figure}
\centerline{\psfig{figure=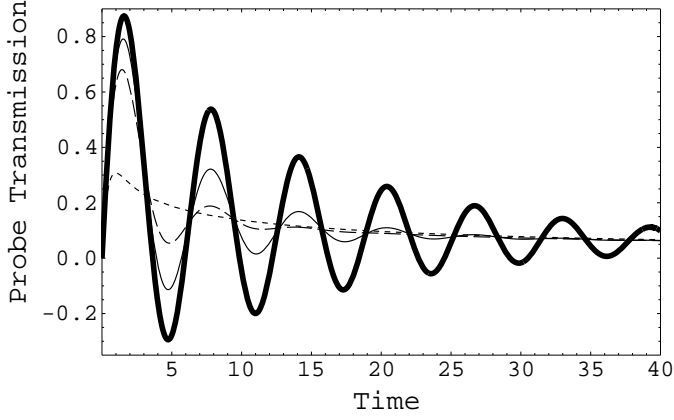,height=5.5cm}}
\caption{\narrowtext  
The time evolution of the imaginary part of the time-dependent 
linear susceptibility
$\left[-\mbox{Im}\left(\chi(t)\right)\right]$ (in arbitrary units).
In our notation positive (negative) values denote probe absorption
(gain). The parameters used were $\delta=\delta_{g}=0$ and
$\gamma=5$ (shot dashed curve), $\gamma=1$ (long dashed curve), 
$\gamma=0.5$ (thin solid curve), and $\gamma = 0.2$
(thick solid curve). Time and $\gamma$ 
are measured in units of $\beta$. }
\label{fig2}
\end{figure}


\begin{figure}
\centerline{\psfig{figure=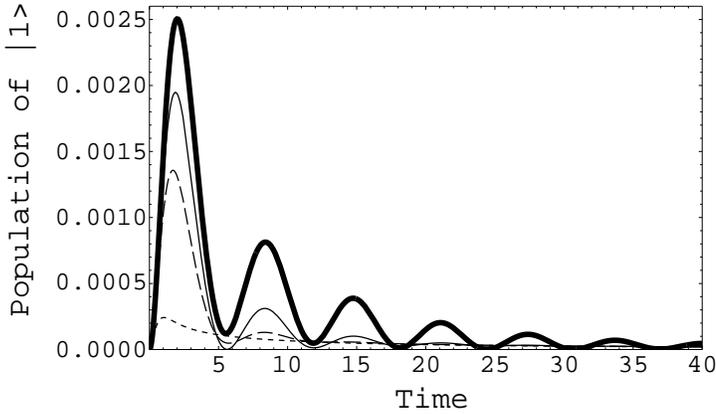,height=5.5cm}}
\caption{\narrowtext  
The time evolution of the population of state $|1\rangle$.
The parameters and the units used are the same as in figure 2. }
\label{fig3}
\end{figure}

\end{document}